\documentclass[twocolumn]{aastex631}
\usepackage{amsmath}
\usepackage{color}
\usepackage{mathrsfs}
\usepackage{natbib}

\defcitealias{krishnaprasad2017}{\scshape KP17}

\shorttitle{MHD Simulations of a Sunspot}
\shortauthors{MacBride et~al.}

\graphicspath{{./}{figures/}}

\begin{document}

\title{Ambipolar Diffusion in the Lower Solar Atmosphere: MHD Simulations of a Sunspot}

\author[0000-0002-9901-8723]{Conor D. MacBride}
\affiliation{Astrophysics Research Centre, School of Mathematics and Physics, Queen’s University Belfast, Belfast, BT7 1NN, UK}

\author[0000-0002-9155-8039]{David B. Jess}
\affiliation{Astrophysics Research Centre, School of Mathematics and Physics, Queen’s University Belfast, Belfast, BT7 1NN, UK}
\affiliation{Department of Physics and Astronomy, California State University Northridge, Northridge, CA 91330, USA}

\author[0000-0003-3812-620X]{Elena Khomenko}
\affiliation{Instituto de Astrof\'{i}sica de Canarias, 38205 La Laguna, Tenerife, Spain}
\affiliation{Departamento de Astrof\'{i}sica, Universidad de La Laguna, 38205 La Laguna, Tenerife, Spain}

\author[0000-0001-5170-9747]{Samuel~D.~T. Grant}
\affiliation{Astrophysics Research Centre, School of Mathematics and Physics, Queen’s University Belfast, Belfast, BT7 1NN, UK}

\correspondingauthor{Conor D. MacBride}
\email{cmacbride01@qub.ac.uk}

\begin{abstract}
	Magnetohydrodynamic (MHD) simulations of the solar atmosphere are often performed under the assumption that the plasma is fully ionized.
	However, in the lower solar atmosphere a reduced temperature often results in only the partial ionization of the plasma.
	The interaction between the decoupled neutral and ionized components of such a partially ionized plasma produces ambipolar diffusion.
	To investigate the role of ambipolar diffusion in propagating wave characteristics in the photosphere and chromosphere, we employ the {\sc Mancha3D} numerical code to model magnetoacoustic waves propagating through the atmosphere immediately above the umbra of a sunspot.
	We solve the non-ideal MHD equations for data-driven perturbations to the magnetostatic equilibrium and the effect of ambipolar diffusion is investigated by varying the simulation to include additional terms in the MHD equations that account for this process.
	Analyzing the energy spectral densities for simulations with/without ambipolar diffusion, we find evidence to suggest that ambipolar diffusion plays a pivotal role in wave characteristics in the weakly ionized low density regions, hence maximizing the local ambipolar diffusion coefficient.
	As a result, we propose that ambipolar diffusion is an important mechanism that requires careful consideration into whether it should be included in simulations, and whether it should be utilized in the analysis and interpretation of particular observations of the lower solar atmosphere.
\end{abstract}

\section{Introduction}
\label{sec:introduction}

It is becoming clear that the degree of plasma ionization in the solar atmosphere must be accounted for in a wide variety of situations.
Plasma with a particularly large number of neutrals often has significantly modified properties, resulting in changes to the physical processes that drive important aspects of solar dynamics \citep{pandey2008,arber2007,ni2015,khomenko2018,khomenko2021,martinez-gomez2021,popescubraileanu2021}, see \citet{ballester2018} for a review.

Within the solar atmosphere, plasma at photospheric and chromospheric heights is partially ionized, with an ionization degree of approximately $10^{-4}-10^{-1}$ \citep{ni2015,vernazza1981}.
This is a result of the photospheric and chromospheric plasma having lower temperatures and gas pressure decreasing with height.
Accounting for the physical processes occurring in both the partially ionized and fully ionized regimes, as well as across their interface, is important for investigating the effect of waves in heating the corona \citep{pandey2008}.

The plasma is composed of three species, ions, electrons and neutrals, interacting via elastic or non-elastic collisions, ion-neutral, electron-neutral, and electron-ion.
All the collision frequencies are a function of temperature and number density of colliding particles, and their values strongly depend on height.
Partial ionization effects primarily arise due to the substantial quantity of neutral species in the plasma, and their insufficient collisional coupling to the  charged species, leading to additional physical processes that need to be taken into consideration.
For example, accounting for the presence of neutrals, and the degree of coupling between the neutral and charged plasma species, significantly affects the Biermann battery effect \citep{biermann1950} and the strength of the magnetic field it generates within the plasma \citep{martinez-gomez2021}.

Understanding the effects of partial ionization in the solar atmosphere is essential for the complete understanding of high-resolution observations.
By comparing the collisional frequencies between the plasma components to the dynamical frequency of the plasma, the level of coupling between the plasma components can be determined.
Based on this level of coupling, particular partial ionization effects become significant.
As highlighted above, partially ionized plasma behaves differently to fully ionized environments, hence employing incorrect plasma conditions in modern numerical simulations may produce unexpected results and render direct comparisons with observations inappropriate.
This can particularly be seen in the study of the dynamics of prominences in the solar atmosphere due to Rayleigh-Taylor and Kelvin-Helmholtz instabilities \citep{ballester2020}.
The abundance of neutrals in the associated plasma has been shown to affect the growth rates and instability thresholds, directly impacting the interpretation of any observation \citep{soler2012,diaz2012,khomenko2014,popescubraileanu2021b,popescubraileanu2021c}.

In this study we focus on the process of ambipolar diffusion, and its significance within the chromospheric layer directly above a sunspot.
Ambipolar diffusion occurs within a partially ionized plasma when the neutrals are not fully coupled to the ionized component.
The ionized component is affected by the Lorentz force, resulting in a magnetic field which is frozen into those plasma species.
However, the decoupled neutral particles do not experience the Lorentz force, and hence the moving ionized particles are in constant collision with the neutrals.
These collisions result in frictional effects between the two components, providing a mechanism for magnetic and mechanical energy to be dissipated, and hence creates a source of localized atmospheric heating \citep{khomenko2018}.

Numerical works by \citet{shelyag2016,popescubraileanu2021} investigated the effect of ambipolar diffusion on chromospheric waves.
While their simulations were constructed to represent the quiet Sun, \citet{popescubraileanu2021} concluded that ambipolar diffusion has a significant damping effect on waves, with areas of greater magnetic field strength exhibiting more localized damping.
Chromospheric waves have also been studied in realistic solar magneto-convection simulations by \citet{khomenko2018}.
They find that, under the influence of ambipolar diffusion, waves in the chromosphere have amplitude variations that depend on both frequency and atmospheric height.
Further to this, \citet{khomenko2021} established that ambipolar diffusion has the effect of dissipating vortex motions that occur over small length scales at atmospheric heights above 200~km.

Previous ambipolar diffusion studies have mainly focused on the quiet Sun over magnetic regions such as sunspot umbrae, where higher magnetic field strengths and lower densities produce a stronger ambipolar diffusion effect on wave energy propagation in the chromosphere \citep{khomenko2014b}.
The strong, aligned magnetic fields above sunspots provide an ideal conduit for wave propagation, allowing energy to be transferred through the solar atmosphere \citep[see, e.g., the review by][]{jess2015}.
The influence of ambipolar diffusion in these regions has already been highlighted by \citet[][henceforth referred to as \citetalias{krishnaprasad2017}]{krishnaprasad2017} in their observations of the mid-photosphere to mid-chromosphere above a sunspot umbra using high-resolution data obtained by the Rapid Oscillations in the Solar Atmosphere \citep[ROSA;][]{jess2010}, the Hydrogen-alpha Rapid Dynamics camera \citep[HARDcam;][]{jess2012}, and the Interface Region Imaging Spectrograph \citep[IRIS;][]{depontieu2014} instruments.
\citetalias{krishnaprasad2017} found that intensity oscillations (which were employed as a proxy for plasma density) in the computed power spectral densities exhibited a power-law dependence at frequencies beyond the dominant $p$-mode peak at $\approx6.5$~mHz.
This power-law dependence was observed across a range of wavelength channels, with the power-law index varying across the different imaging filters (and hence atmospheric heights), suggesting a height-dependency in high-frequency wave attenuation due to the changing physical mechanisms at different heights in the atmosphere.
Furthermore, \citetalias{krishnaprasad2017} highlight that ambipolar diffusion may be a factor in more efficient frequency-dependent damping at higher frequencies.
Such studies highlight that forming a detailed understanding of the effect of ambipolar diffusion in this domain will allow for more precise interpretations of observational results, and allow for more realistic models and simulations of sunspot atmospheres.

In this work, we carry out magnetohydrodynamic (MHD) simulations of magnetoacoustic waves propagating through the atmosphere directly above a sunspot, and analyze the effects that ambipolar diffusion has on the wave properties by comparing simulations with and without ambipolar effects, using identical starting conditions.
In Section~\ref{sec:mancha} we introduce the numerical code employed and in Section~\ref{sec:model} we detail the atmospheric model that was employed in the simulation.
In Section~\ref{sec:results} we outline the results from our numerical experiments and provide further insights and conclusions in Section~\ref{sec:discussion}.

\section{Numerical solution}
\label{sec:mancha}

For this work we are using the \textsc{Mancha3D} code \citep{khomenko2006,felipe2010,khomenko2012b}, which solves the three-dimensional non-linear non-ideal MHD equations for perturbations to the magnetostatic equilibrium.
In particular, \textsc{Mancha3D} can be configured to either include (or exclude) ambipolar diffusion terms in the embedded MHD equations.
When written in their conservative form, the equations \textsc{Mancha3D} is configured to solve in our simulations are,

\begin{equation}
	\dfrac{\partial \rho}{\partial t} + \nabla \cdot \left(\rho \mathbf{v}\right) = 0 \ ,
\end{equation}

\begin{equation}
	\dfrac{\partial{\left(\rho \mathbf{v}\right)}}{\partial t} + \nabla \cdot \left[ \rho \mathbf{v} \mathbf{v} + \left( p + \dfrac{|\mathbf{B}|^2}{2\mu_0} \right)\mathbf{I} - \dfrac{\mathbf{B}\mathbf{B}}{\mu_0} \right] = \rho\mathbf{g} + \mathbf{S}(t) \ ,
\end{equation}

\begin{equation}
	\dfrac{\partial e_\textrm{int}}{\partial t}+ \nabla\cdot \mathbf{v}e_\textrm{int} +
(\gamma -1)e_\textrm{int}\nabla\mathbf{v} =  \eta_A{|\mathbf{J}_\perp|}^2 \ ,
\end{equation}

\begin{equation}
	\dfrac{\partial \mathbf{B}}{\partial t} = \nabla \times \left(\mathbf{v}\times\mathbf{B} - \eta_A\mathbf{J}_{\perp}\right) \ ,
\end{equation}
where $t$ is time, $\rho$ is the density, $\mathbf{v}$ is the velocity, $p$ is the gas pressure, $\mu_0$ is the vacuum permeability, $\mathbf{B}$ is the magnetic field, $\mathbf{I}$ is the identity tensor, $\mathbf{g}$ is the gravitational acceleration, $\mathbf{S}(t)$ is the time-dependent driving force applied during the simulation, and $e_\textrm{int}$ is the internal energy.
In our experiments we used an ideal equation of state (EOS) and therefore $e_\textrm{int}= p/(\gamma -1)$, with the heat capacity ratio $\gamma=5/3$.
The ideal EOS neglects the effects of ionization and molecular dissociation.
The main ionizing species in the chromosphere is hydrogen and its efficient ionization starts after reaching temperatures of about 10~kK \citep{saha1921}.
In our model we only reach temperatures of about 7~kK in the uppermost part of the domain, therefore, the effects of ionization in the EOS should not compromise our results, while allowing for the calculations to be simplified.
Later in the paper it will be shown that the effect of ambipolar diffusion peaks at even lower heights, where the hydrogen stays mostly neutral.
Throughout our numerical calculations we use a constant mean molecular weight of $\mu=1.28$, corresponding to a neutral solar gas mixture.

The current density, $\mathbf{J}$, is given by $\mathbf{J} = \nabla \times \mathbf{B} / \mu_0$, and $\mathbf{J}_\perp$ is the current density in the direction perpendicular to the magnetic field $\mathbf{B}$ and is given by,
\begin{equation}
    \mathbf{J}_\perp = - \dfrac{(\mathbf{J} \times \mathbf{B}) \times \mathbf{B}}{|\mathbf{B}|^2} \ .
\end{equation}
Each of these equations will also contain an additional artificial diffusion term for numerical stability.

The ambipolar diffusion coefficient, $\eta_A$,  is,
\begin{equation}
	\eta_A = \dfrac{\xi^2_n|\mathbf{B}|^2}{\alpha_n} \ ,
\end{equation}
where \(\xi_n\) is the fraction of neutrals, and \(\alpha_n\) is the neutral collisional parameter.
The units of $\eta_A$ are $\mu_0$\,m$^2$\,s$^{-1}$.
For simulations without ambipolar diffusion present, all terms with $\eta_A$ are excluded from the MHD equations employed.

In the MHD equations, the $\eta_A$ terms are always associated with $\mathbf{J}_\perp$ terms.
As a result, ambipolar diffusion only occurs when there is a significant electric current in the direction perpendicular to the magnetic field.
This anisotropy is due to the net diffusion velocity of all plasma components, while only the ions are are affected by the Lorentz force \citep{cowling1957,braginskii1965}.
The ambipolar diffusion coefficient is related to the Cowling resistivity, $\eta_C = \eta + \eta_A$, where $\eta$ is the Ohmic diffusion coefficient.
It is often treated as equivalent when $\eta$ can be ignored, which is the case of the simulations reported here \citep{leake2005,leake2006,arber2007}.

The fraction of neutrals, $\xi_n$, is given by $\xi_n = \rho_n / (\rho_c + \rho_n)$, where $\rho_n$ and $\rho_c$ are the mass densities of neutral and charged components, respectively.
The collisional parameter, $\alpha_n$, is given by,
\begin{equation}
	\alpha_n = \rho_i \nu_{in} + \rho_e \nu_{en} \ ,
\end{equation}
where $\rho_i$ and $\rho_e$ are the mass densities of ions and electrons, and $\nu_{in}$ and $\nu_{en}$ are the collisional frequencies between ions/neutrals and electrons/neutrals, respectively.
The electron, ion, and neutral mass densities are computed by applying the Saha equation, taking account of all the relevant chemical species of the solar atmosphere, namely the first 92 elements from the periodic table with the abundances given in \citet{anders1989}.

The collisional frequencies are \citep{braginskii1965},
\begin{equation}
	\nu_{in} ~=~ n_n \sigma_{in} \sqrt{\dfrac{8 k_B T}{\pi\mu_{in}}} ~\propto~ n_n\sqrt{T} \ ,
\end{equation}
\begin{equation}
	\nu_{en} ~=~ n_n \sigma_{en} \sqrt{\dfrac{8 k_B T}{\pi\mu_{en}}} ~\propto~ n_n\sqrt{T} \ ,
\end{equation}
where the reduced ion/neutral and electron/neutral masses are $\mu_{in} = m_i m_n / (m_i + m_n)$ and $\mu_{en} = m_e m_n / (m_e + m_n)$, respectively, and the ion/neutral and electron/neutral collision cross sections are ${\sigma_{in}=5 \times 10^{-19}}$~m$^2$ and ${\sigma_{en}=1 \times 10^{-19}}$~m$^2$, respectively \citep{huba2013}.

\section{Model and simulation configuration}
\label{sec:model}

The sunspot model is produced following the procedures detailed in \citet{khomenko2008}.
In addition, the model is further modified using the techniques described by \citet{khomenko2009} to make sure that it remains convection-stable.
To ensure this, at large radial distances from the sunspot axis, the model reduces to a quiet Sun model, modified to prevent convective instability.
The standard model `S' from \citet{christensen-dalsgaard1996} is used as the quiet Sun configuration.
However, it is adapted to prevent convective instabilities from developing below the photosphere during the simulation run.
Following the methods documented by \citet{parchevsky2007}, the model is adjusted such that the square of the Brunt-V\"{a}is\"{a}l\"{a} frequencies are positive across all atmospheric heights, hence satisfying the criterion for stability.

\begin{figure}
	\plotone{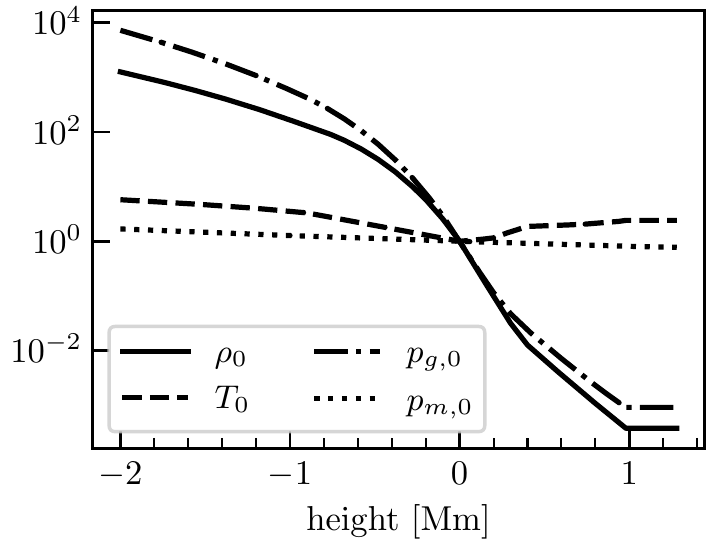}
	\caption{Plot showing the key characteristics of the model. Non-dimensionalized parameters are plotted throughout height at the sunspot axis. The parameters are, density ($\rho_0$), gas pressure ($p_{g,0}$), temperature ($T_0$), and magnetic pressure ($p_{m,0}$). These values have all been normalized to their photospheric value at the temperature minimum on the sunspot axis. Their photospheric values are {$\rho_{\text{ph}} = 4.735 \times 10^{-6}$\,kg\,m$^{-3}$}, {$p_{g,\text{ph}} = 112.4$\,N\,m$^{-2}$}, {$T_{\text{ph}} = 2857$\,K}, and {$p_{m,\text{ph}} = 2758$\,N\,m$^{-2}$}.}
	\label{fig:model}
\end{figure}

To perform 2.5-dimensional (2.5D) wave simulations, our three-dimensional sunspot model first undergoes a coordinate transformation from cylindrical coordinates $(r, \phi, z)$ to Cartesian coordinates $(x, y, z)$.
Then, a vertical slice is taken through the axis of the sunspot, along the $y$-axis, such that the $B_y$ component of the magnetic field is equal to zero.
Some of the key characteristics of the model are shown in Figure~\ref{fig:model}.

The resulting two-dimensional simulation domain is $14.2$\,Mm in the horizontal direction and $2.8$\,Mm in the vertical direction, of which $1$\,Mm is above the photosphere.
The sunspot's axis is located at the center of the simulation domain.
The separation between numerical grid points is $50$\,km in the horizontal direction and $20$\,km in the vertical direction.
For a 2.5D simulation, we place a single grid point along the $y$-axis and set the vector quantities in the MHD equations governing {\sc Mancha3D} to use all three Cartesian directions, though in practice, since $B_y$ is zero in the simulation plane, only perturbations in the $x$- and $z$-components of the velocity and magnetic field vectors will develop.
However, the code also computes perpendicular currents, which in this case will be along the $y$-direction, necessitating a 2.5D setup.

The Courant-Friedrichs-Lewy (CFL) numerical stability condition is used by \textsc{Mancha3D} to determine the time step, which results in the time step in the simulation with ambipolar diffusion typically varying between $0.001 - 0.008$~seconds.
The dynamic time step used by the simulation is the minimum time from the advection time step, the diffusion time step, and a fixed maximum time step provided during initial configuration of the simulation (i.e., $0.008$~seconds).
The advection time step will decrease as the maximum absolute value of the Alfv\'{e}n speed, the sound speed, and the plasma flow velocity increase.
The diffusion time step will generally decrease as the
ambipolar ($\eta_A$), and artificial ($\nu$) diffusion coefficients increase.

\begin{figure}
	\plotone{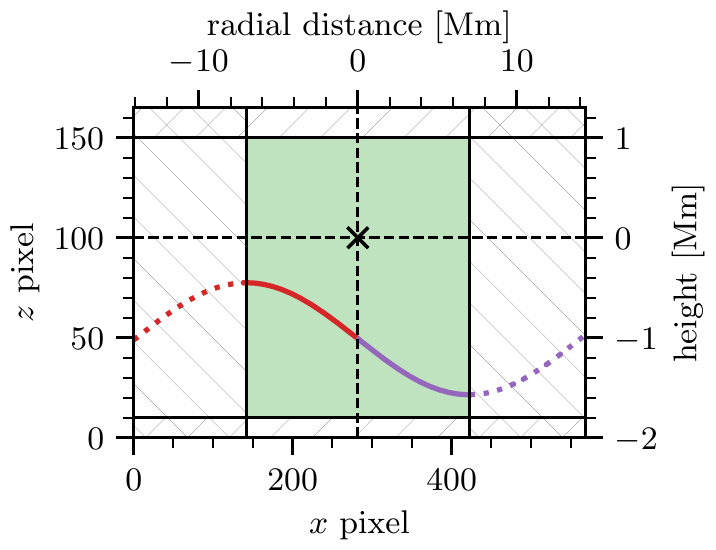}
	\caption{Diagram showing the grid layout of the simulation and the location of key numerical features. The locations along the width of the sunspot model are given in both pixel coordinates (lower) and physical coordinates (upper). Similarly, the locations along the height of the sunspot model are given for pixel coordinates (left) and physical coordinates (right). The vertical dashed line is the axis of the sunspot, while the horizontal dashed line highlights the photospheric height, which is taken to be at the temperature minimum along the sunspot axis. The simulation is driven from the point where these lines intersect, and this point is shown with a cross. The green shaded region is the simulation domain which will be analyzed, while the region outside is only used for numerical purposes. The blue diagonal hatching shows the upper and lower PMLs, while the orange diagonal hatching shows the left and right regions simulating periodic boundary conditions. The red and purple sine curve demonstrates how the model is reflected horizontally to simulate the periodic boundaries, with each dotted line showing how the region with the same color of solid line is reflected horizontally.}
	\label{fig:grid}
\end{figure}

Periodic boundary conditions are used for the left and right (furthest distances away from the central axis of the sunspot umbra) boundaries.
This is implemented by reflecting the model horizontally over the right border and then shifting the model by half the original width to the right such that the sunspot remains in the center of the $x$-axis.
This produces a model with 568 grid points along the horizontal direction, of which 284 grid points are inside the central simulation domain.
The entire numerical domain is large enough such that waves exiting from the left and right boundaries do not re-enter the simulation domain or affect its physics.
For the upper and lower boundaries, perfectly matching layers \citep[PMLs;][]{berenger1994} are used to absorb waves such that open boundaries are simulated.
The upper PML is comprised of 15 additional rows of grid points above the simulation domain (300\,km thick), while the lower PML is comprised of 10 additional rows below (200\,km thick).
Inside the upper PML, gravity has been set to zero to assist with numerical stability.
By removing gravity, waves can propagate upwards more freely, which improves the absorption of perturbations by the PML.
To remain consistent with a zero gravity atmosphere, density ($\rho_0$), temperature ($T_0$), and gas pressure ($p_{g,0}$) must all be constants within the zero gravity region.
As both the MHD equations and the model have been suitably adjusted for a zero gravity atmosphere, the simulation remains in magnetohydrostatic (MHS) equilibrium, and there are no reflections at the PML boundary.
This strategy has been previously applied by \citet{khomenko2012}.
The layout of the simulation is shown in detail in Figure~\ref{fig:grid}.

Perturbations are introduced at the photosphere to drive waves through the atmospheric layers directly above the sunspot.
To best represent the effect of waves in the sunspot atmosphere, we base our wave driver on observationally acquired Doppler velocity information from the Helioseismic and Magnetic Imager \citep[HMI;][]{schou2012} onboard the Solar Dynamics Observatory \citep[SDO;][]{pesnell2012} spacecraft.
The observational data was acquired from 14:36:45 -- 17:36:45~UT on 2014 August 30, with the observations reprojected to account for differential solar rotation using algorithms available in SunPy \citep{sunpy_community2020,mumford2022}.
These data observe the same sunspot (part of active region NOAA~12149) that was observationally studied by \citetalias{krishnaprasad2017}.

Vector magnetic field information from the HMI/SDO instrument was calculated using the Very Fast Inversion of the Stokes Vector \citep[VFISV;][]{borrero2011} algorithm, enabling the inclination angles of the sunspot magnetic field to be extracted.
The umbral pixel with the most vertical magnetic field inclination was identified and a cross-cut through this location was used to extract the one-dimensional Doppler velocity driver.
To ensure the resulting sunspot wave driver is devoid of long-term trends (e.g., to account for solar rotation affecting the component of velocity directed along the spacecraft's line of sight), a mean quiet Sun velocity time series was extracted, fitted with a low-order polynomial, and the resulting trend was subtracted from the one-dimensional sunspot wave driver.
A low-frequency sine function was then subtracted from the driver such that the perturbations oscillate around zero, ensuring the time series is stationary.
The driver was also multiplied by a constant coefficient of $10^{-3}$ to reduce the maximum perturbation value, improving numerical stability, resulting in a driver with an RMS force density of $0.0481$~N\,m$^{-3}$.
Finally, the driver is spline interpolated to decrease the cadence of the observed velocities from 45~seconds down to 1~second.

\begin{figure}
	\plotone{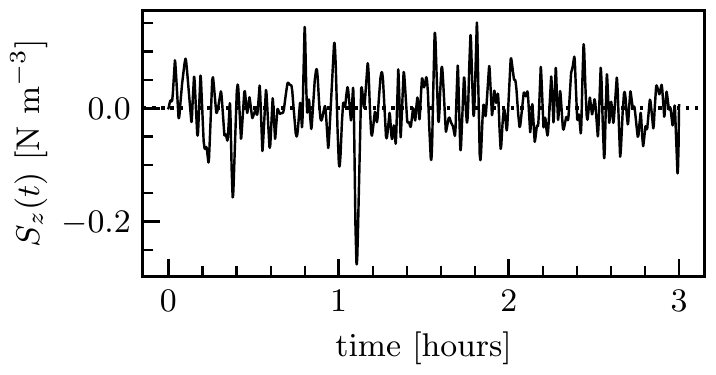}
	\caption{The observationally obtained driving force, $\mathbf{S}(t) = S_z(t) \hat{\mathbf{z}}$, applied to the sunspot atmosphere during the simulation around the sunspot's central axis at photospheric heights. The HMI/SDO Doppler velocity observations this driver is based upon have been processed as outlined in Section~\ref{sec:model}. The horizontal dotted line represents a zero driving force.}
	\label{fig:driver}
\end{figure}

This one-dimensional driver is applied as a perturbation with units of force density in the MHD equations, as the $z$-component of $\mathbf{S}(t)$, at the base of the photosphere on the axis of the sunspot, with a Gaussian reduction to a zero perturbation around this point using a sigma of 3 grid points in both directions.
As the driver has a cadence of 1~second and \textsc{Mancha3D} utilizes a much shorter time step, within \textsc{Mancha3D}, linear interpolation is applied between successive driver values.
In order to prevent numerical issues associated with applying a strong driving force to an atmosphere initially at rest, at the start of the simulation the driving force is multiplied by a Gaussian, $1-\exp{\left(-t^2/(2\sigma^2)\right)}$, where $\sigma = 50$~seconds.
This has the effect of gradually increasing the force from zero at the start of the simulation.
The final value of $S_z(t)$ is shown in Figure~\ref{fig:driver}.

\section{Results and Discussion}
\label{sec:results}

\begin{figure*}
    \plotone{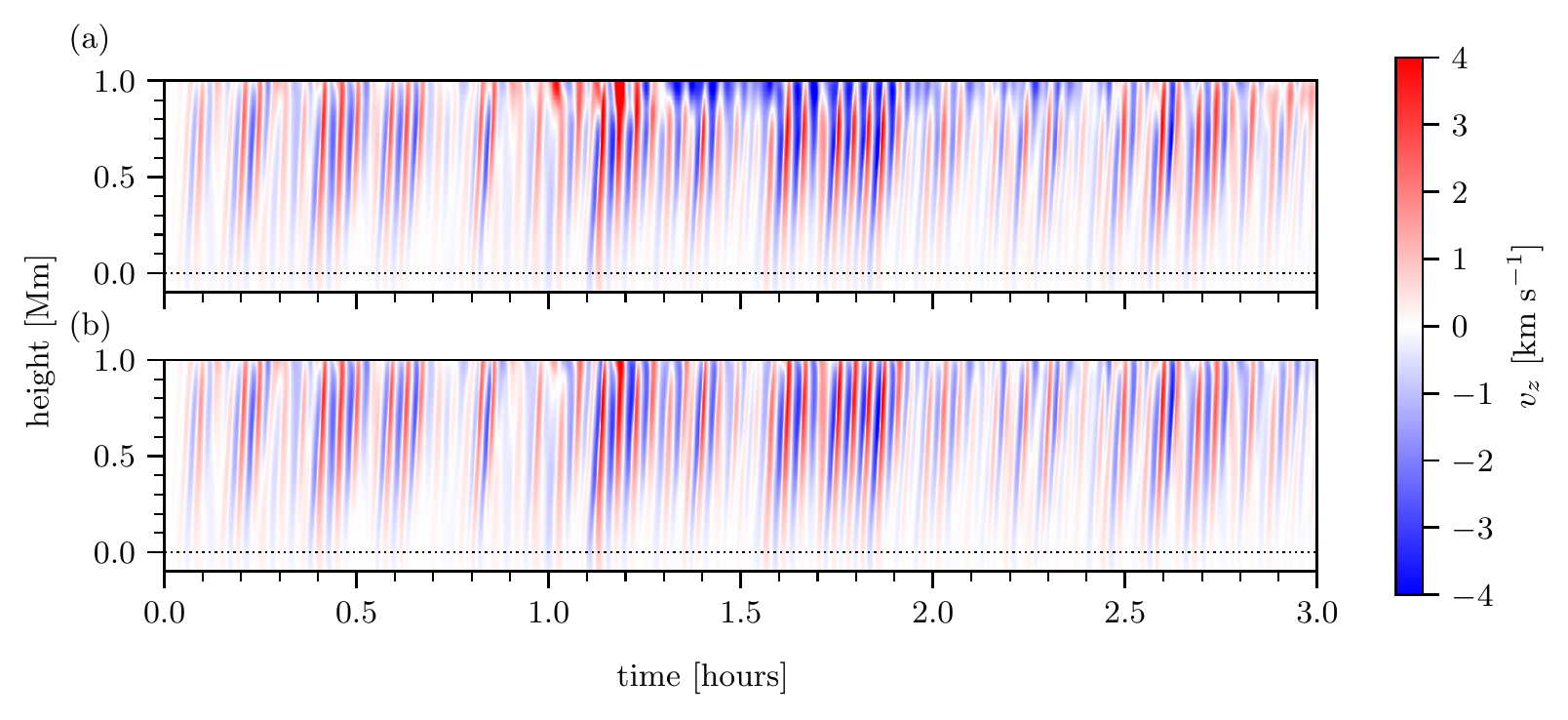}
    \plotone{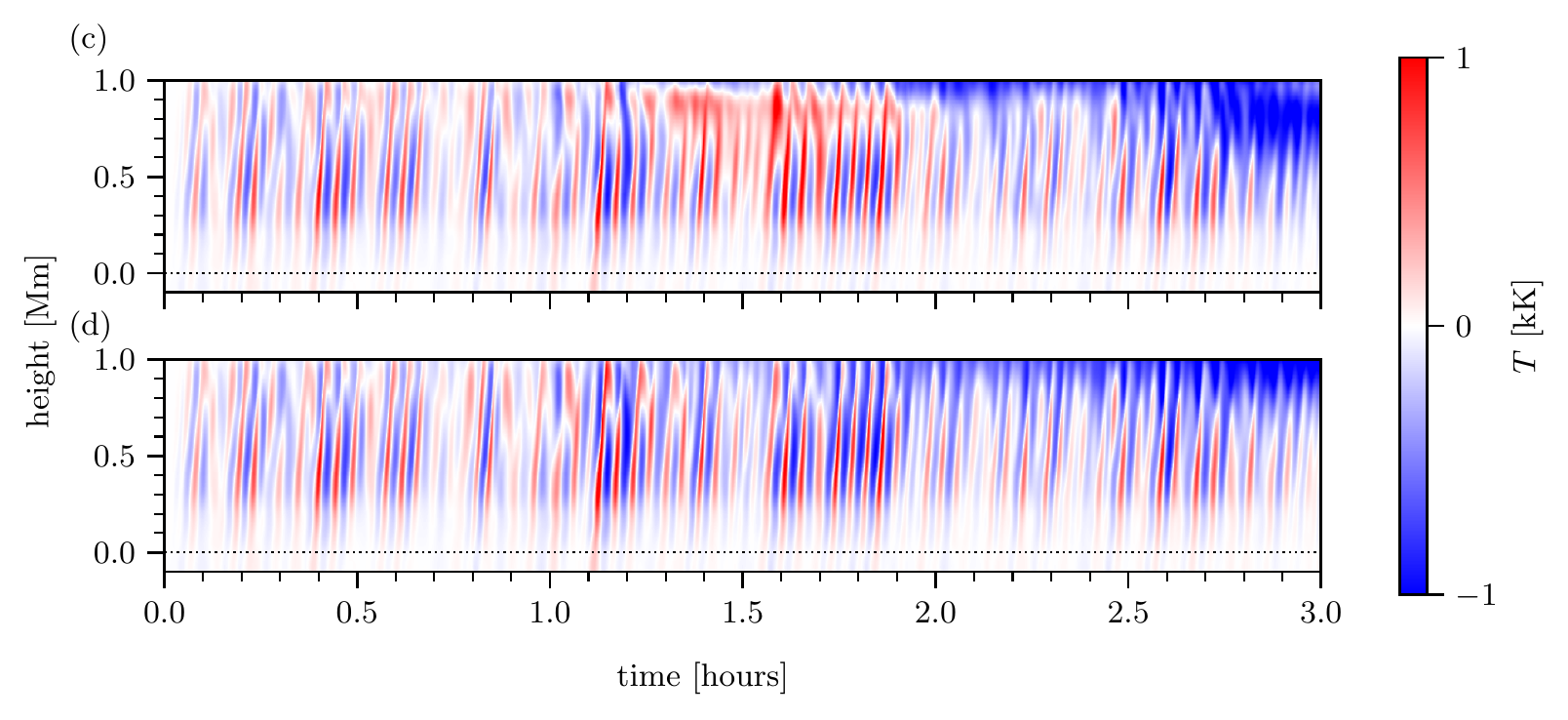}
	\caption{Plots of parameter perturbations during the simulation throughout time ($x$-axis) and atmospheric height ($z$-axis) along the axis of the sunspot. The $z$-component of velocity perturbations, $v_z$, are shown in (a) and (b), and the temperature perturbations, $T$, are shown in (c) and (d) for simulations with (panels a \& c) and without (panels b \& d) ambipolar diffusion. Horizontal dotted lines show the location of the driver.}
	\label{fig:paramtime}
\end{figure*}

\subsection{Simulated parameters}

Two simulations are evaluated using our initial sunspot model across 3~hours of solar evolution time, with one simulation modeling ambipolar diffusion, whilst the other simulation has the ambipolar diffusion terms excluded.
Snapshots of the plasma parameters are saved every 10~seconds of simulation time, resulting in a total of 1080 snapshots per simulation.

As a tool to monitor the simulation, we can inspect how the plasma parameters vary throughout time at significant locations in the numerical domain, such as along the axis of the sunspot.
Figure~\ref{fig:paramtime} shows the value of the $z$-component of velocity, $v_z$, and the temperature, $T$, for perturbations throughout time and atmospheric height along the sunspot axis, hence creating a time-distance diagram for each plasma parameter.
In Figure~{\ref{fig:paramtime}}, time-distance diagrams with (panels a \& c) and without (panels b \& d) ambipolar diffusion are shown for comparison.
These plots also identify the photospheric layer where the wave-driving perturbations are positioned.

\subsection{Energy spectral density}
In order to investigate the frequency dependence of ambipolar diffusion as a function of geometric height, we calculate the autospectral density of the simulated velocity-time series at each atmospheric height output by the {\sc Mancha3D} code.
As can been seen in Figure~\ref{fig:paramtime}, the output velocity data, $v_z$, is non-stationary, i.e., its statistical properties vary throughout time \citep{bendat2010}.
Therefore, we must calculate energy spectra rather than power spectra.

The energy autospectral density function, $\mathscr{G}_{xx}(f)$, is related to the power autospectral density function, $G_{xx}(f)$, as,
\begin{equation}
	\mathscr{G}_{xx}(f) = L G_{xx}(f) \ ,
\end{equation}
where $L$ is the length of data available and $G_{xx}(f)$ is the one-sided autospectral density function given by,
\begin{equation}
	G_{xx}(f) = 2 \lim_{L \rightarrow \infty} \dfrac{1}{L} E\Big[ |X(f)|^2 \Big] \ ,
\end{equation}
where $X(f)$ is the finite Fourier transform of the velocity-time series with a sampling interval of $L$ seconds and $E[\dots]$ represents the expected value \citep{bendat2010}.
As we are working with a discrete time series, we apply the Fast Fourier Transform (FFT) algorithm and also use an estimate of $G_{xx}(f)$,
\begin{equation}
	\tilde{G}_{xx}(F) = \dfrac{2}{L} |X(f)|^2 \ .
\end{equation}
This results in the one-sided energy autospectral density, which we call the energy spectrum, namely \citep{stull1988},
\begin{equation}
	\mathscr{G}_{xx}(f) = 2 |X(f)|^2 \ .
\end{equation}

We apply this equation independently to the time series extracted at each pixel along the sunspot's height axis ($z$-axis in the numerical domain).
This allows for the frequency response of waves to be compared across atmospheric heights.
When applying the FFT algorithm to the $z$-component of velocity, $v_z(t)$, we exclude the initial 1000~seconds of the simulation in order to give the driven oscillations time to strengthen.
The maximum frequency analyzed is limited by the 45~second cadence of the underlying HMI/SDO observations, providing us with a Nyquist frequency of $\approx11.1$~mHz.
To aid with the comparison to other studies, we divide all energy spectra by the corresponding frequency sampling, $\Delta f = 1/L = 1 / (10\,800~\textrm{s} - 1000~\textrm{s}) \approx 0.102~\textrm{mHz}$, to provide an energy spectral density (i.e., in units of km$^{2}$\,s$^{-2}$\,mHz$^{-1}$) that is suitable for scientific examination.

\begin{figure}
\gridline{\fig{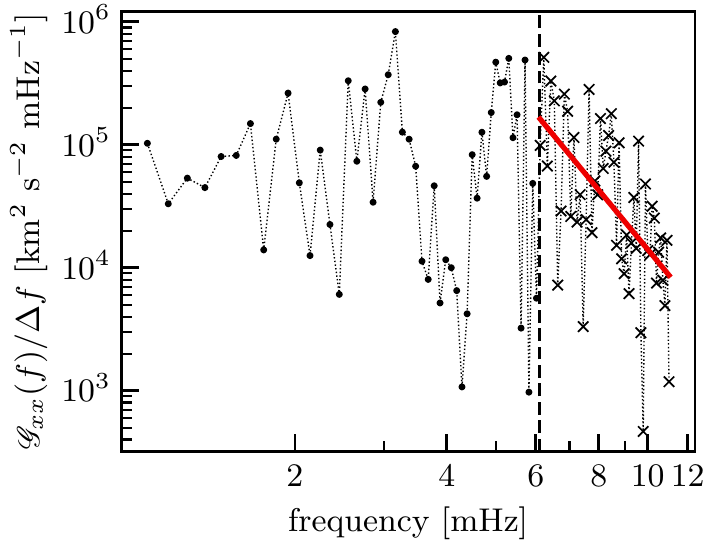}{0.9\columnwidth}{(a)}}
\gridline{\fig{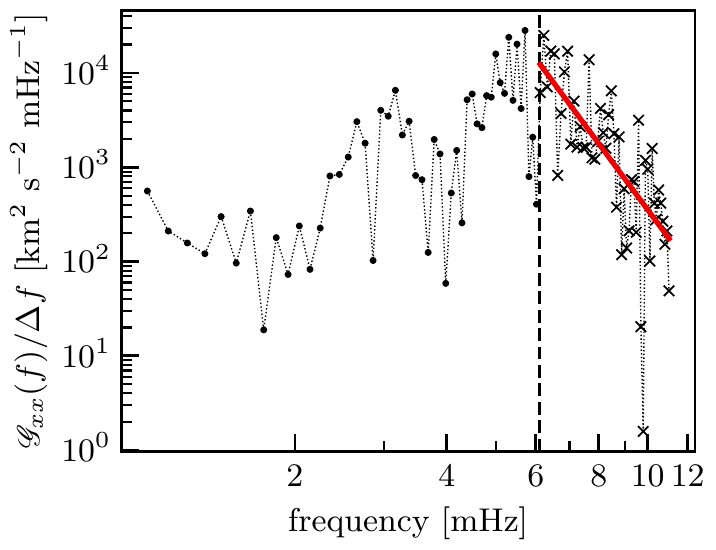}{0.9\columnwidth}{(b)}}
	\caption{Plots of the energy spectral density at different heights within the solar atmosphere along the $z$-axis of the sunspot for the simulation with ambipolar diffusion terms included in the MHD equations. The top of the simulation domain ($\approx1$~Mm) is shown in (a), while the location of the driver ($\approx0$~Mm) is shown in (b). The vertical dashed line is at $\approx6.1$~mHz, after which the spectral points are marked with crosses instead of dots. A power-law fit is applied to the points with frequencies $>6.1$~mHz, where the fit is shown with a solid red line.}
	\label{fig:esd_ad}
\end{figure}

\begin{figure}
\gridline{\fig{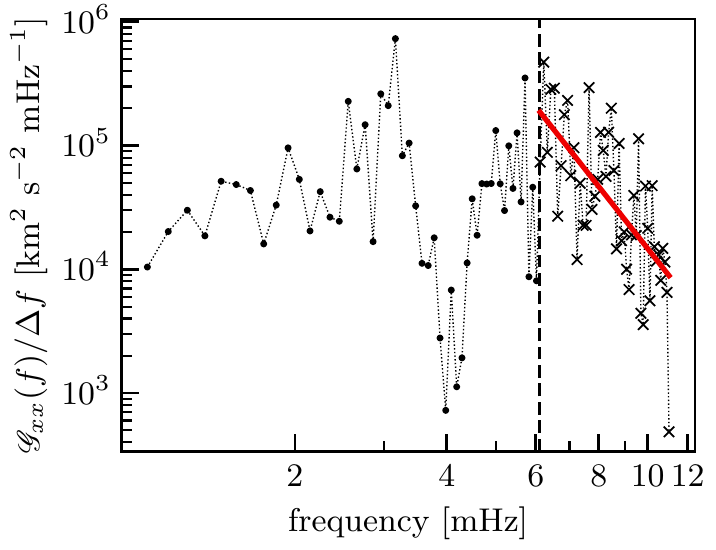}{0.9\columnwidth}{(a)}}
\gridline{\fig{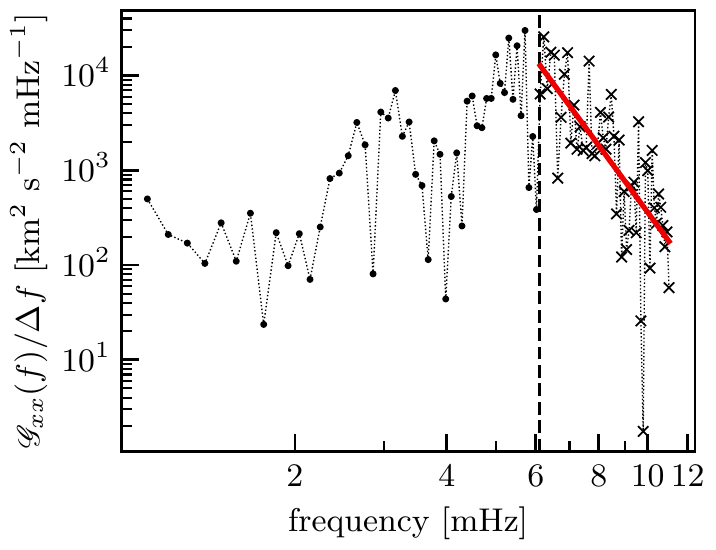}{0.9\columnwidth}{(b)}}
	\caption{Identical plots to those shown in Figure~{\ref{fig:esd_ad}}, only now with no ambipolar diffusion included in the relevant MHD equations.}
	\label{fig:esd_nad}
\end{figure}

Furthermore, by calculating the energy spectral densities for both simulations independently, this allows us to infer how wave behavior changes with atmospheric height, and hence compare how the presence of ambipolar diffusion affects these characteristics.
Figures~\ref{fig:esd_ad} \& \ref{fig:esd_nad} show examples of energy spectral densities plotted for a number of key atmospheric heights across both simulations.
At the location of the driver the energy spectral densities between both simulations have similar values, while at the top of the simulation domain the energy spectral densities exhibit more distinct differences between the two simulations.


\subsection{Power-law indices}
As highlighted in the work of \citetalias{krishnaprasad2017}, observations of a sunspot umbra acquired through different filters (corresponding to different heights in the solar atmosphere) produced varying power-law indices for frequencies above the peak value.
In the study documented by \citetalias{krishnaprasad2017}, the frequency that displayed dominant spectral energy was $\approx6.5$~mHz (corresponding to a period of 152~seconds).
This frequency is consistent with the typical upper-photospheric/chromospheric $p$-mode spectrum, where only frequencies greater than the acoustic cutoff frequency (i.e., $>5$~mHz) are permitted to propagate \citep[see, e.g.,][]{kanoh2016,murawski2016,felipe2018b}.
For frequencies higher than this dominant value, \citetalias{krishnaprasad2017} found power-law slopes of the energy spectral densities to vary depending on the layer of the atmosphere being sampled.
In our present work, Figures~{\ref{fig:esd_ad}} \& {\ref{fig:esd_nad}} reveal a spectral peak at $\approx6.1$~mHz (periodicity of 164~seconds), which is consistent with the observational work of \citetalias{krishnaprasad2017}.
At higher frequencies (i.e., $>6.1$~mHz), we find the energy spectral densities decrease progressively up to the Nyquist frequency of $\approx11.1$~mHz.

By measuring the slopes of the energy spectral densities beyond the $\approx6.1$~mHz enhancement, we are able to calculate a power-law index for this frequency region at each atmospheric height, for both simulations, in order to compare the effects of ambipolar diffusion on the resulting power-law slopes.
The energy spectral densities with a frequency beyond the enhancement are selected, with a logarithmic function applied to the frequencies and energy spectral densities such that the linear relationship can be determined.
An unweighted least squares first order polynomial fit is applied to the log-log frequencies and energy spectral densities, with the power-law slope taken to be the slope of the fitted first order polynomial.
Examples of the fitted power-law slopes are plotted for each simulation, at specific atmospheric heights, in Figures~\ref{fig:esd_ad} \& \ref{fig:esd_nad}, with the corresponding power-law indices displayed as a function of atmospheric height in Figure~\ref{fig:powlaw}.
The work of \citetalias{krishnaprasad2017} focused on observations spanning the mid-photosphere (e.g., data acquired in the Mg~{\sc{i}}~b$_2$ spectral line) through to the mid-chromosphere (e.g., H$\alpha$ and Ca~{\sc{ii}}~K observations).
As a result, we focus our analysis between the atmospheric heights of $500-1000$~km in order to examine the layers associated with the photospheric/chromospheric interface.

In Figure~{\ref{fig:powlaw}} it can be seen that the power-law index for the simulation without ambipolar diffusion increases monotonically with atmospheric height throughout the $500-1000$~km range.
This is not the case when ambipolar diffusion is accounted for in the simulation, resulting in the power-law index decreasing with atmospheric height between the heights of $680-840$~km in the atmosphere, before increasing again above this height, as seen in Figure~\ref{fig:powlaw}.

\subsection{Ambipolar diffusion coefficient}
To understand the differences between the power-law slopes arising in the simulations with/without ambipolar diffusion, we can analyze and decompose the ambipolar diffusion coefficient, $\eta_A$, in the MHD equations.
As discussed in Section~\ref{sec:mancha}, the coefficient is given by $\eta_A = \xi^2_n|\mathbf{B}|^2/\alpha_n$.
The magnetic field, $|\mathbf{B}|$, and the temperature, $T$, do not vary significantly between the range of atmospheric heights spanning $500-1000$~km, so their effects on the power-law index differences within this range can be neglected.

\begin{figure}
	\plotone{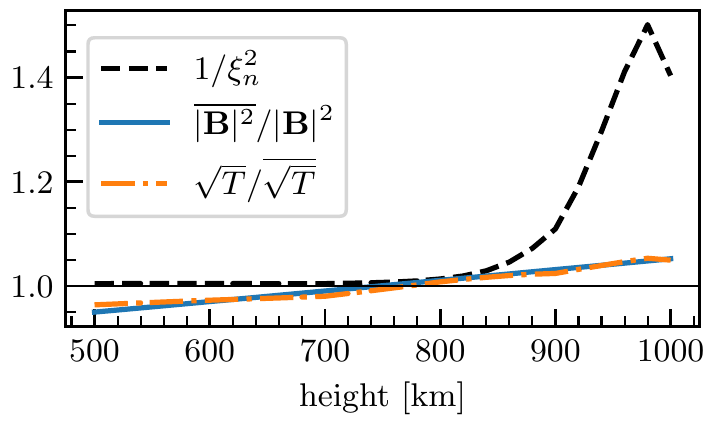}
	\caption{Plots of the individual approximations applied to the ambipolar diffusion coefficient, $\eta_A$, to produce $\eta_A'$ as shown in Equation~\ref{eq:eta_a_approx}. The approximation $1 / \xi_n^2$ is plotted using a dashed black line, $\overline{|\mathbf{B}|^2} / |\mathbf{B}|^2$ is plotted using a solid blue line, and $\sqrt{T} / \overline{\sqrt{T}}$ is plotted with a dash-dotted orange line. The solid black horizontal line highlights the value of these approximations in relation to unity. All of the terms are extracted from the {\sc Mancha3D} simulation with ambipolar diffusion along the $z$-axis of the sunspot. The mean of the terms are then taken throughout time to get their average value at each atmospheric height. The terms with an overline use the mean value calculated over all dimensions.}
	\label{fig:eta_terms_approx}
\end{figure}

In particular, the magnetic field squared, $|\mathbf{B}|^2$, within the simulation with ambipolar diffusion varies approximately linearly from $0.0062$~T$^2$ at 500~km to $0.0056$~T$^2$ at 1000~km, and has a standard deviation of $0.00018$~T$^2$.
As a relatively constant value is maintained, the mean magnetic field, $\overline{|\mathbf{B}|^2} = 0.0059$~T$^2$, is used in the analysis.
This approximation is applied by multiplying the ambipolar diffusion coefficient, $\eta_A$, by $\overline{|\mathbf{B}|^2} / |\mathbf{B}|^2$, which can be seen in Figure~\ref{fig:eta_terms_approx} to be relatively close to unity.

In Section~\ref{sec:mancha}, we introduced the neutral collisional parameter, $\alpha_n = \rho_i \nu_{in} + \rho_e \nu_{en}$.
We also showed that the collisional frequencies are proportional to $n_n \sqrt{T}$.
Therefore, the neutral collisional parameter can be re-written as $\alpha_n \propto n_e n_n \sqrt{T}$.
Furthermore, the ambipolar diffusion term can be examined analytically, where it can be shown that,
\begin{align}
	\eta_A & \propto \xi_n^2 \dfrac{1}{\alpha_n} \\
	& \propto \dfrac{n_n^2}{(n_e+n_n)^2} \dfrac{1}{n_e n_n \sqrt{T}} \\
	& \propto \dfrac{n_n}{n_e(n_e+n_n)^2} \dfrac{1}{\sqrt{T}} \ .
\end{align}

\begin{figure}[!t]
	\plotone{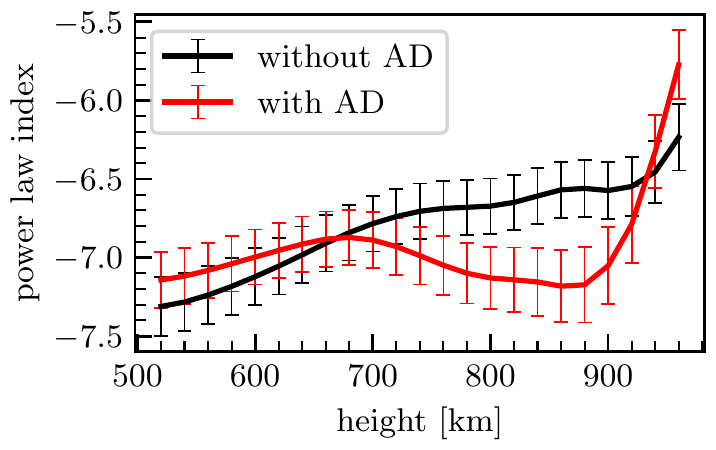}
	\caption{Plot of the power-law indices of the energy spectral densities within the frequency interval of $6.1-11.1$~mHz, calculated at different heights within the solar atmosphere along the $z$-axis of the sunspot. The black line is for the simulation without ambipolar diffusion (AD), while the red line is for the simulation that contains AD terms in the associated MHD equations.}
	\label{fig:powlaw}
\end{figure}

In the case where there is a high degree of plasma ionization (i.e., $n_e \gg n_n$), the ambipolar diffusion coefficient tends towards zero as expected.
However, for a partially ionized plasma with $n_n \approx n_e$ (i.e., $\xi_n \approx \frac{1}{2}$) or $n_n \gg n_e$ (i.e., $\xi_n \approx 1$), the equation follows that,
\begin{align}
	\eta_A & \propto \dfrac{1}{\alpha_n}\label{eqn:coefficient} \\
	& \propto \dfrac{1}{n_e n_n \sqrt{T}} \ .
\end{align}

\begin{figure}
	\plotone{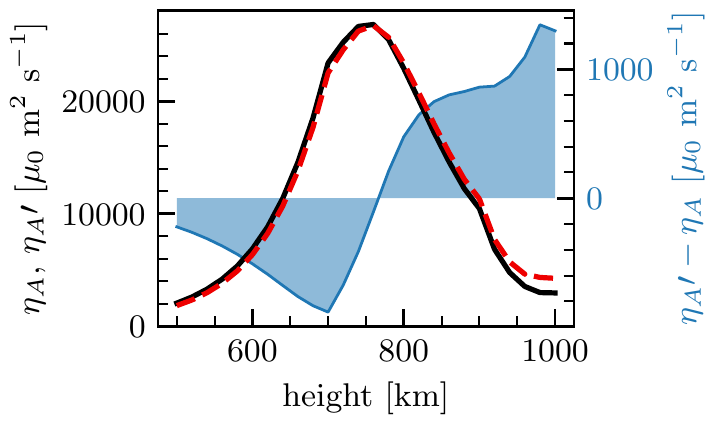}
	\caption{Plots of the ambipolar diffusion coefficient, $\eta_A$, along with its approximation, $\eta_A'$. The coefficient $\eta_A$ is plotted using a solid black line against the left axis, while $\eta_A'$ is plotted using a dashed red line against the left axis. The difference between both lines, $\eta_A' - \eta_A$, is plotted using a solid blue line against the right axis. All of the terms are extracted from the {\sc Mancha3D} simulation with ambipolar diffusion along the $z$-axis of the sunspot. The mean of the terms are then taken throughout time to get their average value at each atmospheric height.}
	\label{fig:eta_terms}
\end{figure}

For the partially ionized plasma in our sunspot model, this results in a proportionality between the ambipolar diffusion coefficient, $\eta_A$, and collisional parameter, $\alpha_n$.
Therefore, we would expect ambipolar diffusion to have the most significant effect on partially ionized plasmas when the collisional parameter is particularly small.
In the case where the temperature does not vary significantly, this corresponds to plasma where the product of the number densities is small, which is the case in the ambipolar diffusion simulation presented in this study.
Relative to the variation in the product of the number densities, temperature is approximately constant over the height range of $500-1000$~km and the relationship simplifies to $\alpha_n \propto n_e n_n$.
In particular, the square root of the temperature varies from a minimum of 67~K$^{1/2}$ to a maximum of 86~K$^{1/2}$, with a mean of 78~K$^{1/2}$ and a standard deviation of 3~K$^{1/2}$, which is an insignificant variation compared to variations of the number densities.
The product of the number densities, $n_e n_n$, varies from a minimum at $10^{33}$ to a maximum at $10^{37}$, with a mean value of $2 \times 10^{35}$ and a standard deviation of $4 \times 10^{35}$.
The mean temperature is therefore used in the analysis, with the ambipolar diffusion coefficient, $\eta_A$, adjusted by multiplying it by $\sqrt{T} / \overline{\sqrt{T}}$.
Figure~\ref{fig:eta_terms_approx} shows that the effect of this adjustment on $\eta_A$ is small.
This final adjustment leads to,
\begin{equation}
	\eta_A \propto \dfrac{1}{n_e n_n} \ .
\end{equation}
The combined effect of all these approximations is shown in Figure~\ref{fig:eta_terms} where the ambipolar diffusion coefficient, $\eta_A$, is plotted alongside its approximation,
\begin{equation}
	\eta_A' = \eta_A
	\dfrac{\overline{|\mathbf{B}|^2}}{|\mathbf{B}|^2}
	\dfrac{1}{\xi_n^2} \dfrac{\sqrt{T}}{\overline{\sqrt{T}}}\ .\label{eq:eta_a_approx}
\end{equation}

Comparing these analytical results to outputs provided by the {\sc Mancha3D} simulation, we find close agreement with our theoretical considerations above.
Figures~{\ref{fig:eta_terms_approx} \& \ref{fig:eta_terms}} highlight that it is the collisional parameter, $\alpha_n$, rather than the neutral fraction, $\xi_n$, that is significant in determining the value of the ambipolar diffusion coefficient, $\eta_A$, as analytically shown in Equation~{\ref{eqn:coefficient}}.
Furthermore, comparing Figure~\ref{fig:eta_terms} to Figure~\ref{fig:powlaw}, we are able to see that the reciprocal of the product of the number densities, i.e., $\eta_A \propto 1/(n_e n_n)$, becomes significantly large only within the range of atmospheric heights where the greatest difference in power-law index is observed between the simulations with and without ambipolar diffusion.
Therefore, this implies that the ambipolar diffusion coefficient, $\eta_A$, which is proportional to the reciprocal of the collisional parameter, $\alpha_n$, is responsible for the steepening of the power-law index between the atmospheric heights of $680-840$~km.

The steeper downward power-law slopes suggest that ambipolar diffusion leads to increased damping of higher frequency waves.
\citet{khomenko2018} show this effect occurring in simulations of magneto-convection within the lower solar atmosphere.
The authors find that when ambipolar diffusion is modeled, the power at higher frequencies is diminished to a greater extent than their lower frequency counterparts.
The observational work of \citetalias{krishnaprasad2017} also suggests that ambipolar diffusion may enhance frequency-dependent damping, which is now readily visualized in Figure~{\ref{fig:powlaw}}.

\begin{figure}
	\plotone{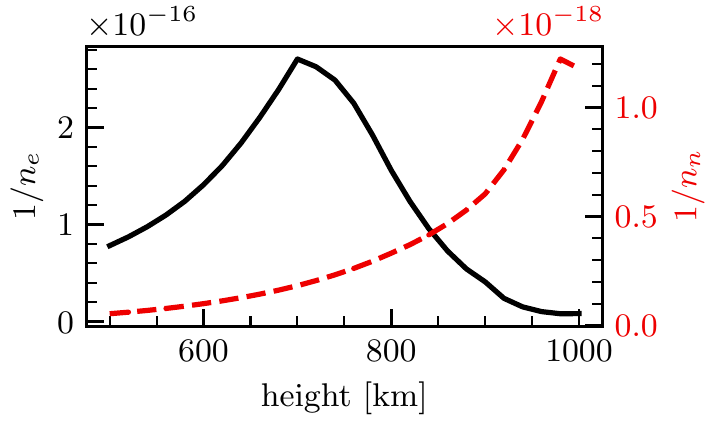}
	\caption{Plots of the reciprocals of the number densities for electrons ($1/n_e$) and neutrals ($1/n_n$). The electron number density reciprocal is represented by a solid black line against the left axis, whilst the the neutral number density reciprocals plotted using a dashed red line against the right axis. The reciprocals are calculated at every point along the sunspot $z$-axis and the time-average value is calculated.}
	\label{fig:collisional}
\end{figure}

In Figure~\ref{fig:collisional}, we plot the reciprocals of the terms $n_e$ and $n_n$ separately.
On average, the reciprocal of the electron number density ($1/n_e$) has a bell-shaped curve with a peak at 700~km, the height where there are fewest free electrons.
On the other hand, the reciprocal of the neutral number density ($1/n_n$) is small initially, but increases gradually as atmospheric height increases, meaning that, on average, there are fewer neutrals higher in the atmosphere.
It is these two terms multiplied together that results in the neutral collisional term, $1/\alpha_n$, exhibiting the bell-shaped curve that is shown in Figure~\ref{fig:eta_terms}, which, on average, has a peak value at 760~km in the atmosphere.
As a result, we are able to conclude that the maximal interplay between the reciprocals of the $n_e$ and $n_n$ terms (maximizing the atmospheric distributions when there are fewest free electrons and the smallest number of neutrals --- i.e., corresponding to the least dense plasma) provides the strongest effect on the wave power-law indices.
These effects are clearly visible in Figures~{\ref{fig:powlaw}}, {\ref{fig:eta_terms}} \& {\ref{fig:collisional}}, where strong deviations in the corresponding plots are evident at an atmospheric height of $\sim760$~km.

\section{Conclusions}
\label{sec:discussion}

Magnetohydrodynamic wave simulations were evaluated in 2.5D for a cross-section of the chromosphere above a sunspot umbra, with the energy spectral densities calculated at each atmospheric height.
Two simulations were produced independently, one with ambipolar diffusion terms included in the embedded MHD equations and one without.
We compare the energy spectral densities between the two simulations, focusing our analysis within the frequency range spanning $6.1-11.1$~mHz, which can be fitted with a power-law relationship following the observational work of \citetalias{krishnaprasad2017}.
We focus our analysis on the atmospheric heights spanning $500-1000$~km, where previous wave studies of the mid-photosphere to mid-chromosphere have been performed on the same sunspot structure \citepalias{krishnaprasad2017}.

From the simulations, we provided evidence that increases in the ambipolar diffusion coefficient, $\eta_A$, was a result of the collisional parameter, $\alpha_n$, becoming smaller.
We showed that in our sunspot model the collisional parameter is proportional to $n_e n_n$, where $n_e$ and $n_n$ are the number densities of the electrons and neutrals, respectively.
Using this relationship, we are now able to state that the process of ambipolar diffusion in a sunspot atmosphere is significant, and needs to be accounted for, when the value of $1/(n_e n_n)$ becomes relatively large.
We find the value of the ambipolar diffusion coefficient ($\eta_A \propto 1/n_e n_n$) to be closely linked with variations seen between the two simulations presented in this study.
At atmospheric heights where there was an insignificant difference between the energy spectral density power-law slopes of both simulations, the ambipolar diffusion coefficient remained relatively low.
This resulted in the terms modeling ambipolar diffusion to have an insignificant effect on the plasma parameters.
However, at atmospheric heights where significant variations were present in the power-law slopes of the energy spectral densities, the ambipolar diffusion coefficient was found to be particularly large.
Under this regime (approximately $680-920$~km above the solar surface), ambipolar diffusion caused the power-law slopes in the range of $6.1-11.1$~mHz to become steeper, hinting at increased wave energy damping in this environment.
Interestingly, we do not find that the ionization fraction, $\xi_e$, has a significant role in determining where ambipolar diffusion is important in the sunspot atmosphere, even though it is part of the ambipolar diffusion coefficient, $\eta_A$.

The importance of the product of the electron number density, $n_e$, and the neutral number density, $n_n$, highlights that ambipolar diffusion is more significant in less dense plasma.
While in general we can expect ambipolar diffusion to be more important higher in the atmosphere where the plasma is less dense, we also need to consider the relative ionization of each of the plasma components.

This work provides evidence of ambipolar diffusion having a significant effect on the plasma properties within an umbral sunspot atmosphere and provides a number of insights for future sunspot studies.
The inclusion of ambipolar diffusion terms in MHD simulations of sunspot atmospheres should be fully considered, particularly when the ambipolar diffusion coefficient ($\eta_A$) approximation ($1 / n_e n_n$) is large.
We also suggest that ambipolar diffusion may produce a clear effect on the power-law indices of energy spectral densities that is detectable in observations of the lower solar atmosphere of sunspots.

The next stage of this work will be to acquire spectral observations \citepalias[to obtain Doppler velocity measurements instead of the intensity signatures employed by][]{krishnaprasad2017} of a sunspot across a range of wavelengths and analyze how the energy spectral density power-law indices vary over atmospheric height.
We would expect to observe a steepening of the power-law slopes at similar atmospheric heights to our simulation when ambipolar diffusion was included ($680-840$~km as shown in Figure~\ref{fig:powlaw}).
This would provide direct observational evidence of ambipolar diffusion occurring within the lower chromosphere above a sunspot.

Next generation spectral imaging instruments, such as the Visible Tunable Filter \citep[VTF;][]{schmidt2016} and the Diffraction Limited Near Infrared Spectropolarimeter (DL-NIRSP), which will be available on the Daniel K. Inouye Solar Telescope \citep[DKIST;][]{tritschler2016}), will be essential for this work.
High-precision spectral observations over a fine spatial and temporal grid are vital to calculate robust power-law slopes of energy spectral densities to compare to the slopes from our simulations.
In addition to the observational instruments, advanced software tools will be required to extract precise Doppler velocity measurements from the spectral images \citep{macbride2021}, which can be combined with cutting-edge inversion routines \citep[e.g.,][]{beck2019,delacruzrodriguez2019,ruizcobo2022} to provide better height resolution of where the Doppler signatures are formed, hence improving the reliability of comparisons between observations and theory.

	C.D.M. would like to thank the Northern Ireland Department for the Economy for the award of a PhD studentship.
	D.B.J. is grateful to Invest NI and Randox Laboratories Ltd. for the award of a Research \& Development Grant (059RDEN-1).  D.B.J. also wishes to thank the UK Space Agency for a National Space Technology Programme (NSTP) Technology for Space Science award (SSc~009).
	D.B.J and S.D.T.G thanks the UK Science and Technology Facilities Council (STFC) for the consolidated grant ST/T00021X/1.
	E.K. thanks the support by the European Research Council through the Consolidator Grant ERC-2017-CoG-771310-PI2FA and by the Spanish Ministry of Economy and the Industry and Competitiveness through the grant PGC2018-095832-B-I00.
	Data supporting this study are available upon request from the corresponding author.

\bibliography{ManchaPaper}

\end{document}